\documentclass[12pt,preprint]{aastex}

\usepackage{natbib}

\newcommand\xmmc{2XMM J011942.7+032421}%
\newcommand\radec{(R.A.,~decl)}%
\newcommand\msun{$M_{\odot}$}%
\newcommand\av{$A_{\rm V}$}%
\newcommand\kms{$\rm km\;s^{\rm -1}$}%
\newcommand\ergs{$\rm ergs\;s^{\rm -1}$}%
\newcommand{\hi}{\ion{H}{1}}
\newcommand{\hii}{\ion{H}{2}}
\newcommand{\hei}{\ion{He}{1}}
\newcommand{\heii}{\ion{He}{2}}
\newcommand{\nitrogeni}{[\ion{N}{1}]}
\newcommand{\nii}{[\ion{N}{2}]}

\newcommand{\neiii}{[\ion{Ne}{3}]}
\newcommand{\sii}{[\ion{S}{2}]}
\newcommand{\oi}{[\ion{O}{1}]}
\newcommand{\oii}{[\ion{O}{2}]}
\newcommand{\oiii}{[\ion{O}{3}]}

\newcommand{\ariii}{[\ion{Ar}{3}]}
\def\simgt{\lower.5ex\hbox{$\; \buildrel > \over \sim \;$}}
\def\simlt{\lower.5ex\hbox{$\; \buildrel < \over \sim \;$}}

\shorttitle{Optical study of \xmmc}
\shortauthors{Guti\'errez, C. M., \& Moon, D.-S.}
\slugcomment{Accepted for publication in Astrophysical Journal  Letters}

\begin{document}

\title{Optical study  of the hyper-luminous X-ray source\\ \xmmc}

\author{Carlos. M. Guti\'errez$^{1,2}$, and Dae-Sik Moon$^{3}$}
\affil{$^1$Instituto de Astrof\'\i sica de Canarias, E-38205 La Laguna, Tenerife,
SPAIN\\ $^2$Departamento de Astrof\'\i sica, Universidad de la Laguna, E-38206
Tenerife, SPAIN\\
$^3$ Department of Astronomy and Astrophysics, University of Toronto, Toronto, ON M5S 3H4, Canada\\
}
\email{cgc@iac.es}

\begin{abstract}
We present the identification and characterization of the optical counterpart 
to 2XMM J011942.7+032421, one of the most luminous and distant ultra-luminous 
X-ray sources (ULXs). The counterpart is located near a star forming region 
in a spiral arm of the galaxy NGC~470 with $u$, $g$, and $r$ magnitudes of
21.53, 21.69, and 21.71 mags, respectively. The luminosity of the counterpart
is much larger than that of a single O-type star, indicating that it may be a stellar cluster.
Our optical spectroscopic observations confirm the association of the X-ray
source and the optical counterpart with its host galaxy NGC~470,
which validates the high, $\simgt$ 10$^{41}$ \ergs, X-ray luminosity of the source.
Its optical spectrum is embedded with numerous emission lines, including
H recombination lines, metallic forbidden lines and more notably the high-ionization
\heii\ ($\lambda$4686 \AA) line.
This line shows a large velocity dispersion of $\simeq$ 410 \kms, 
consistent with the existence of a compact ($<$ 5 AU) highly-ionized accretion disc
rotating around the central X-ray source.
The $\sim$ 1.4 $\times$ 10$^{37}$ \ergs\ luminosity of the \heii\ line emission 
makes the source one of the most luminous ULXs in 
the emission of that line. This, together with the high X-ray luminosity and the 
large velocity dispersion of the \heii\ emission, suggests that the source
is an ideal candidate for more extensive follow-up observations for 
understanding the nature of hyper-luminous X-ray sources,
a more luminous subgroup of ULXs and more likely candidates for intermediate-mass black
holes.
\end{abstract}


\section{Introduction}

Stellar black holes remnants of massive stars have masses of up to $\sim$ 30 \msun,
typically radiating at luminosities of $\sim$ 10$^{37}$ \ergs\ and occasionally up to  
$\sim$ 10$^{39}$ \ergs\ \citep[see][]{mc06}. 
Supermassive black holes at the centers of galaxies, on the other hand,
have typical masses in the range of 10$^6$--10$^9$ \msun\ and luminosities of  
$10^{41}$--$10^{45}$ \ergs.
Between the two lie putative intermediate-mass black holes (IMBHs)
of 10$^2$--10$^4$ \msun\ proposed to explain the high X-ray luminosities of 
10$^{39}$--10$^{42}$ \ergs\ from ultra-luminous X-ray sources (ULXs),
which are bright off-nuclear X-ray point sources mostly found near 
star-forming regions in spiral galaxies \citep[e.g.,][]{lb05}.
The IMBHs interpretation of the ULXs is particularly compelling for the more luminous subset
of ULXs of $\simgt$ 10$^{41}$ \ergs\ for which other explanations for their nature,
such as those based on beamed emission or supper-Eddington accretion onto compact objects \citep[e.g.,][]{om11,me01,met03},
have hitherto been unable to provide plausible scenarios. 
This is consistent with the recent results from X-ray population studies \citep{set11}
which separate such luminous, i.e., $\simgt$ 10$^{41}$ \ergs, ULXs from 
the population of stellar-mass X-ray sources 
in the local Universe, bolstering the case for the IMBH interpretation.

Although ULXs have been studied mostly in the X-ray waveband,
identification and characterization of ULXs in the optical,
where one can study their stellar counterpart, accretion disc, 
and/or extended nebular emission -- 
particularly the high-ionization 
\heii\ ($\lambda$4686 \AA; 54 eV ionization potential) emission 
from X-ray photo-ionization --
is essential to understanding the nature of these sources 
\citep[e.g.,][]{vdM04,gut06,gut13,lg06,met11}. 
Here, we present a study on 2XMM J011942.7+032421 \citep{set12,gla13},
one of the most luminous ULXs
located at $\sim$ 33\arcsec\ (or $\sim$ 5.3 kpc at 33.5 Mpc)
from the center of the SA(rs)b galaxy NGC 470,
in the optical waveband.
Located at the host galaxy NGC 470 as confirmed in this study (see \S2),
the maximum X-ray luminosity in the 0.2--12 keV range of the ULX is 
$\sim$ 1.53 $\times$ 10$^{41}$ \ergs\ \citep{set12}, and
its X-ray spectra can be fit either by absorbed power-law emission or 
absorbed multicolor-disk (MCD) emission of $\sim$ 1~keV inner disc temperature
with its mass of $\simgt$ 1,900 \msun.
The source has exhibited an order of magnitude decrease of its X-ray luminosity 
in power-law emission, 
and there is some evidence for the existence of a short-term, i.e., $\simlt$ 1 hr, variability,
indicating accretion onto a massive compact object.
The host galaxy NGC~470 forms a pair with NGC~474 at $\sim$ 5\arcmin\ away, and
the pair shows clear signs for early interactions between them \citep{ret06}.

The high, $\simgt$ 10$^{41}$ \ergs, X-ray luminosity of \xmmc\ places it 
in the aforementioned group of more luminous ULXs, dubbed ``hyper-luminous'' 
(or sometimes ``extreme'') ULXs \citep[e.g.,][]{soret10, get13}.
Among them only two other sources of M82 X-1 and ESO 243-49 HLX-1
have been studied with their counterparts in the optical waveband to the best of our knowledge.
M82 X-1 was identified to be associated with a stellar cluster \citep{pzet04},
but a recent study showed that the position of the optical counterpart 
may have a significant offset, $\simgt$ 0\farcs65 (or 3 $\sigma$ level),
from its X-ray position \citep{vet11}. 
The optical counterpart to ESO 243-49 HLX X-1 has an intrinsic brightness 
$M_R$ $\simeq$ --11, comparable to that of a massive globular cluster \citep{soret10},
and was detected with H$\alpha$ emission from a redshift comparable to that of the host
galaxy ESO 243-49 \citep{wet10}. 
Although the detection of the H$\alpha$ emission confirms the location of HLX X-1 
in the host galaxy, no detailed optical spectroscopic information is avaiable 
for this source or M82 X-1.
Our study presented in this $Letter$ of \xmmc\ is the first detailed spectroscopic
observations of a hyper-luminous ULX in the optical waveband.

\section{Observations and Results}

Our inspection of SDSS images and catalogues\footnote{http://skyserver.sdss3.org/dr10/en/home.aspx}
at the location of the ULX \xmmc\
reveals the presence of a point source (ID \#1237678620110880953) at 
\radec\ = ($\rm 01^h19^m42\fs{77}$, $\rm +03\arcdeg24\arcmin22\farcs{6}$) (J2000)
near the Chandra X-ray position 
($\rm 01^h19^m42\fs{8}$, $\rm +03\arcdeg24\arcmin22\arcsec$)
of the ULX reported in \citet{set12}.
(Note that no archival HST image for the region around the ULX is available.)
We measure its more precise X-ray position using the same Chandra data in Sutton et al. to be 
\radec\ = 
($\rm 01^h19^m42\fs{76}$, $\rm +03\arcdeg24\arcmin22\farcs{3}$).
This is consistent with the previous measurement 
and is $\sim$ 0\farcs3 away from the optical point source in the SDSS data,
which is well within the positional uncertainty resident in matching the Chandra and SDSS astrometry.
The $u$, $g$, and $r$ magnitudes of the optical point source
are 21.53 $\pm$ 0.13, 21.69 $\pm$ 0.06 and 21.71 $\pm$ 0.08 mag, 
respectively, while its $i$ and $z$ magnitudes have rather large uncertainties 
and are excluded from further analyses (see below).
Based on the position and also on the spectroscopic information 
presented below, we identify this object
to be the optical counterpart to the ULX 2XMM J011942.7+032421.

Spectroscopic observations of the optical counterpart were made with
the OSIRIS spectrograph\footnote{http://www.gtc.iac.es/instruments/osiris/osiris.php}
on the 10.4-m GTC\footnote{http://www.gtc.iac.es}
in La Palma (Spain) using its service observing mode on 2012 September 14.
Three consecutive 900-s long-slit spectra of R1000B grism were obtained 
at the parallactic angle of each exposure under $\sim$ 1\arcsec\ seeing 
condition around $\sim$ 1.2 airmass. 
The slit width and length were 1\arcsec\ and 8\arcmin, respectively; 
the pixel spatial sampling was 0\farcs25.
The  spectra were analyzed following the standard procedure using 
IRAF\footnote{IRAF is the Image Reduction and Analysis Facility, 
written and supported by the IRAF programming group at the National Optical Astronomy
Observatories (NOAO) in Tucson, Arizona.}, which comprises bias
subtraction, flat-field correction, co-addition of the three single exposures,
wavelength calibration, and spectral extraction. The flat-field correction was
made only in the red ($\ge$ 6,300 \AA) part of each spectrum due to the low response of the
calibration lamp in the blue part. 
Spectral calibration was conducted using 32 He-Ar and Ne arc lines that we obtained
at the beginning of the night in the wavelength range $\lambda$ $\simeq$ 4,000--7,500 \AA,
and the rms uncertainty in the wavelength calibration was $\simeq$ 0.06 \AA.
The FWHM of the lines, 
which is equivalent to the intrinsic spectral resolution of the OSIRIS spectrograph, 
increases from $\simeq$~6.0 \AA\ at 4,000 \AA\ to $\simeq$~7.6 \AA\ at 7,500 \AA.
The spectroscopic standard star Ross 640 (Oke 1990) was observed for 
photometric calibration.

\begin{figure*}
\includegraphics[angle=0,scale=.60]{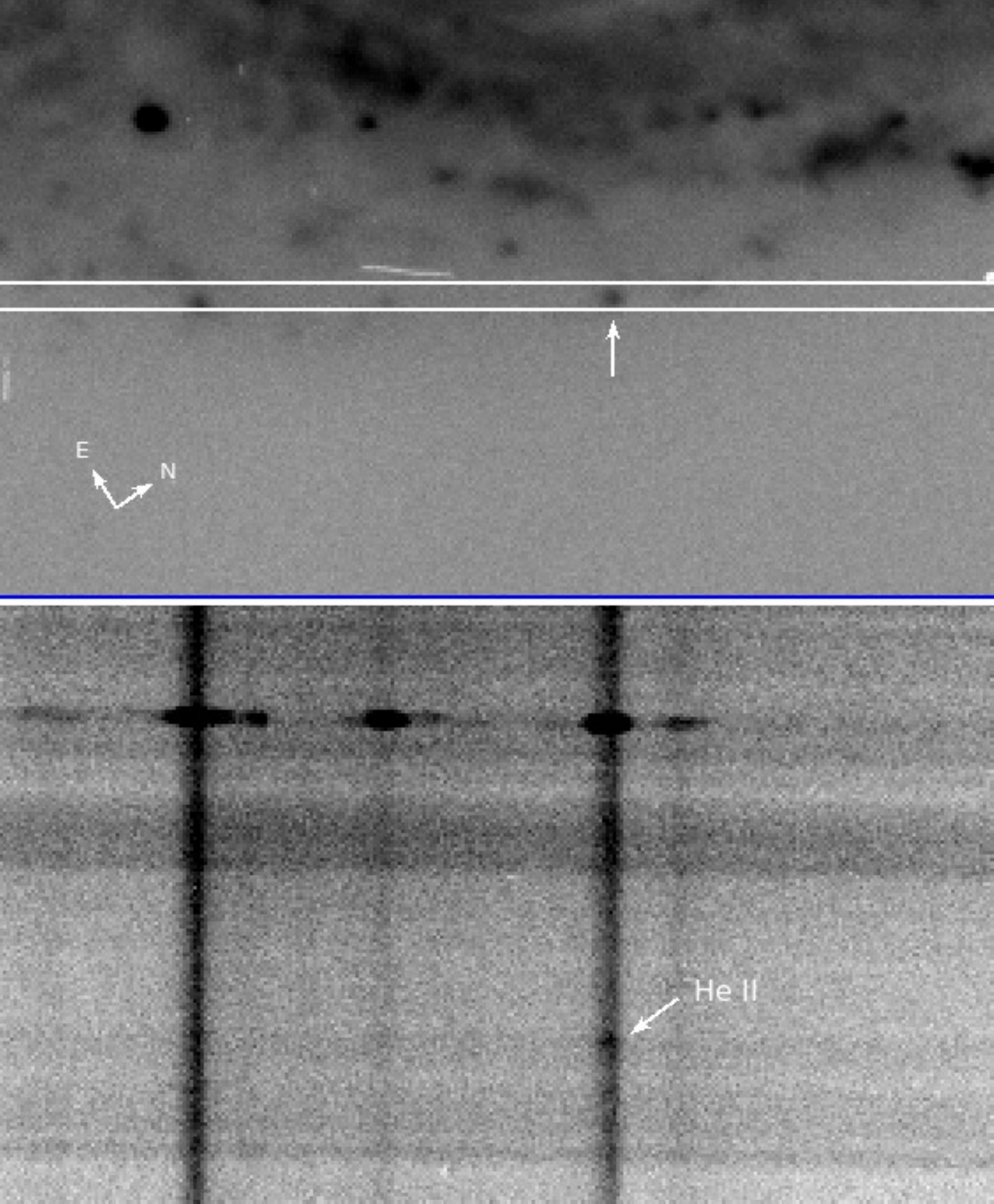}
\caption{($Top$) Optical image of the region around the position of \xmmc. 
The optical counterpart to the source is indicated by the small white vertical arrow. 
The two thick white horizontal lines mark the position and width (1\arcsec) of the slit used to obtain the optical spectrum. 
($Bottom$) A section of the 2-D dispersed image containing the wavelengths of \heii\ ($\lambda$4686 \AA) and 
H$\beta$ emission. The small arrow marks the location of the \heii\ emission from the source.}
\label{fig:disp}
\end{figure*}

The top panel in Figure~\ref{fig:disp} is a raw 20-s acquisition image of the field obtained by ORSRIS
just prior to taking spectra, showing the location of the source, which is indicated by a white arrow, 
overlaid on the slit projection;
the bottom panel shows part of a 2-D dispersed image of the obtained OSIRIS spectrum including
the wavelength range of H$\beta$ (4861 \AA; extended emission around continuum sources) 
and \heii\ (4868 \AA; indicated by a white arrow) lines.
As in Figure~\ref{fig:disp}, in addition to \xmmc, spectra of three more sources 
lying on the slit were obtained: two at $\sim$ 29\farcs4 and 16\farcs2
away from \xmmc\ in the left-hand side and one at $\sim$ 5\farcs90 away 
in the right-hand side.
All the three sources show extended H recombination lines with relative velocities 
of $\sim$ --78, --38, and --19 \kms, respectively, to \xmmc.
Note that only \xmmc\ shows the \heii\ emission line at 4686 \AA.
\citet{hk83} identified 51 \hii\ regions in spiral arms of NGC~470,
18 of them within 1\arcmin\ distance from \xmmc\ including the \hii\ region 46 
inside the positional uncertainty of the ULX.
It is, therefore, highly likely that all the four sources 
in Figure~\ref{fig:disp} are \hii\ regions 
or at least associated with \hii\ regions in spiral arms of the host galaxy NGC~470.

According to \citet{set12}, the best-fit X-ray spectra of \xmmc\ give H column  density
of 1.0--1.4 $\times$ 10$^{21}$ cm$^{-2}$ for absorbed power-law model and $<$ 9 $\times$
10$^{20}$ for absorbed multi-component disk blackbody model in the host galaxy,
additionally to 3.1 $\times$ 10$^{20}$ cm$^{-2}$ in the Galaxy toward the source.  We
take the low limit of 1 $\times$ 10$^{21}$ cm$^{-2}$ from the absorbed power-law
model fits  as the H column density toward the source, and this gives \av\ $\simeq$ 0.70
mag inclusive of the Galactic contribution. This estimation agrees with the extinction
\av\ $=0.56\pm 0.21$ deduced from the empirical H$\alpha$/H$\beta$ ratio \citep{cal00} 
where the quoted uncertainty includes contributions from those in line intensity measurement
and dust models.

Figure~\ref{fig:spectrum} shows the spectrum of the counterpart \xmmc\ which is
wavelength rest-framed and also extinction corrected.
The spectrum is dominated by numerous emission lines, 
including H Balmer series, He lines (both \hei\ and \heii) and forbidden 
lines of heavy elements, 
which overlie blue continuum emission.
Table~\ref{tbl_line} contains a list of the emission lines identified in our spectrum.
The velocity of H$\alpha$ line is $cz_{\rm helio}$ $\simeq$ 2,370 \kms, and it is 
consistent with the value $2,374\pm 1$ \kms\ obtained as the radial velocity 
of NGC~470 in \hi\ observations \citep[e.g.,][]{huc99}.
The rms velocity shift of other lines with respect to the H$\beta$ emission 
is 7.2 $\pm$ 14.8 \kms\ which is insignificantly small. 
No meaningful velocity offset between the \heii\ ($\lambda$4686 \AA) line 
and the other lines is identified as the former shows a --9.8 $\pm$ 5.3 \kms\
shift from the H$\beta$ line,
confirming that the ULX counterpart is indeed located in the host galaxy NGC 470
and, consequently, validating the X-ray luminosity measurement of the ULX (see \S1).
The 9.0 $\pm$ 0.6 \AA\ FWHM of the \heii\ line is significantly 
larger than those of the other isolated lines which have an average FWHM of 6.4 $\pm$ 0.9~\AA.
In order to investigate the large linewidth of the \heii\ line further, 
we measure it in all of the three individual 900-s exposures to be 
11.9 $\pm$ 3.7, 13.0 $\pm$ 2.7 and 16.0 $\pm$ 5.2 \AA, 
consistently larger than those of the other lines.
The increased uncertainties caused by the faintness of the signal in the individual exposures,
however, make it impossible to check if the central velocity of the \heii\ line changes
during the three exposures.
Note that all the isolated lines other than the \heii\ line have a FWHM
compatible with the intrinsic spectral resolution of the OSIRIS spectrograph (see above).
Excluding the instrumental line width of 6.3 \AA\ measured in an arc line at 4,400 \AA,
the FWHM of the \heii\ line corresponds to velocity dispersion of $\simeq$ 410 \kms.
The de-reddend line intensity ratios $\simeq$ 85 and $\simeq$ 1.4 for 
(\nii(6548 \AA)+\nii(6583 \AA))/\nii(5755 \AA) and \sii(6716 \AA)/\sii(6713 \AA), respectively,
indicate the temperature and electron density of
$\simeq$ 1.1 $\times$ 10$^4$ K and $\simeq$ 30 cm$^{-3}$, respectively \citep{of06},
whereas $\simeq$ 11.5 for the ratio of 
(\oii((3727 \AA)+\oiii(4959+5007\AA))/H$\beta$ 
does subsolar metallicity of the source \citep{kd02}.

We use the continuum emission in Figure~\ref{fig:spectrum} to obtain its $g$ and $r$ magnitudes
of 21.8 and 22.1 mag, respectively, after removing the line emission in the bands.
The $g$ band magnitude is comparable to that from the SDSS photometry, 
whereas the $r$ band magnitude is larger than the SDSS result.
We attribute this discrepancy in the $r$ band magnitudes to the contribution by the
strong H$\alpha$ emission of the source and also to the uncertainty resident in
photometric calibration of spectroscopic observations.
The de-reddened absolute magnitudes of the optical counterpart in the $g$ and $r$ bands are 
--11.7 and --11.1 mag, respectively, after correction for its \av\ = 0.7 mag extinction.
The absolute magnitudes of the optical counterpart of \xmmc\
are approximately 100 times brighter 
than a single O-type star \citep{z90} and are comparable to those of a more evolved star
such as a Wolf-Rayet (WR) star or a luminous blue variable \citep[e.g.,][]{het14}, or
more likely an unresolved stellar cluster.

\section{Discussion and Conclusions}

As described in \S2, \xmmc\ is located close to an \hii\ region in NGC~470 with 
small velocity offsets from other nearby \hii\ regions in spiral arms --
which is indicative that it is related to star formation activities therein --
and also from the nucleus of the host galaxy. 
This makes it highly unlikely that the source is a product of
disruption of a satellite galaxy by the host galaxy or a merge of satellite galaxies \citep{bet10}
as suggested for the hyper-luminous ULX ESO 243-49 showing $\ga$ 400 \kms\ velocity shift
from its nucleus \citep{soret12}.
The absence of significant dynamic motions other than the large velocity dispersion in
the  \heii\ emission also makes it unlikely that the source 
is a recoiling supermassive black hole \citep{kzl08}.
And its $\sim$ 300 X-ray (0.5--8 keV range) to optical ($R$ band) flux ratio,
computed by the method in \citet{het01}, 
is much larger than the typical value, i.e., $\simlt$ 10, found
in active galactic nuclei \citep{bet04}, while it is lower
than the value, $\simgt$ 500, found in ESO 243-49 \citep{soret10},
supporting the interpretation that the source is unlikely associated
with a supermassive black hole.

\begin{figure*}
\includegraphics[angle=0,scale=.60]{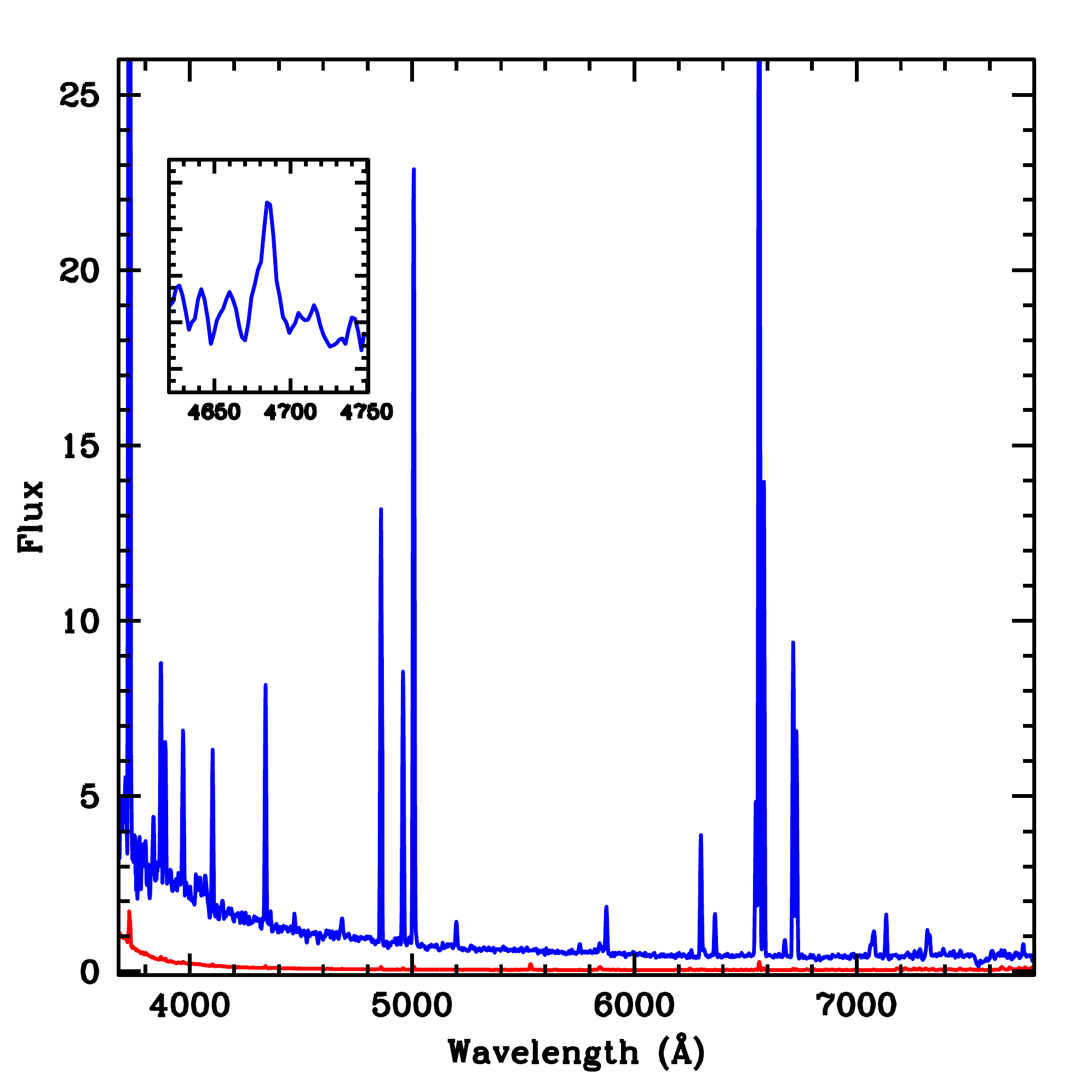} 
\caption{Our OSIRIS spectrum of the optical counterpart to \xmmc.
The flux in the y-axis represents the monochromatic line intensity 
in the unit of 10$^{-17}$ erg s$^{-1}$ cm$^{-2}$ \AA$^{-1}$, and 
the red line shows a background spectrum from the nearby sky.
The inset enhances the \heii\ emisison at 4,686 \AA.}
\label{fig:spectrum}
\end{figure*}

As observed in Cygnus X-3 and NGC 300 X-1 \citep{let05, bet11}, 
WR stars accreting onto a compact object have shown high X-ray luminosities
up to $\sim$ 10$^{38}$ \ergs,
and \citet{let13} suggested that M101 ULX-1 is a binary system of a WR star 
and a black hole of 20--30 \msun.
However, it is less likely that the optical counterpart of \xmmc\ is 
a WR star accreting onto a compact object
since we do not see any spectral signature, e.g., broad emission lines indicating significant mass loss,
of a WR star (Figure~\ref{fig:spectrum}). 
Other possibilities for the origin of the optical continuum emission of \xmmc\
include an unresolved stellar cluster or irradiate disc emission as often found in other
ULXs.

\begin{figure*}
\includegraphics[angle=0,scale=.60]{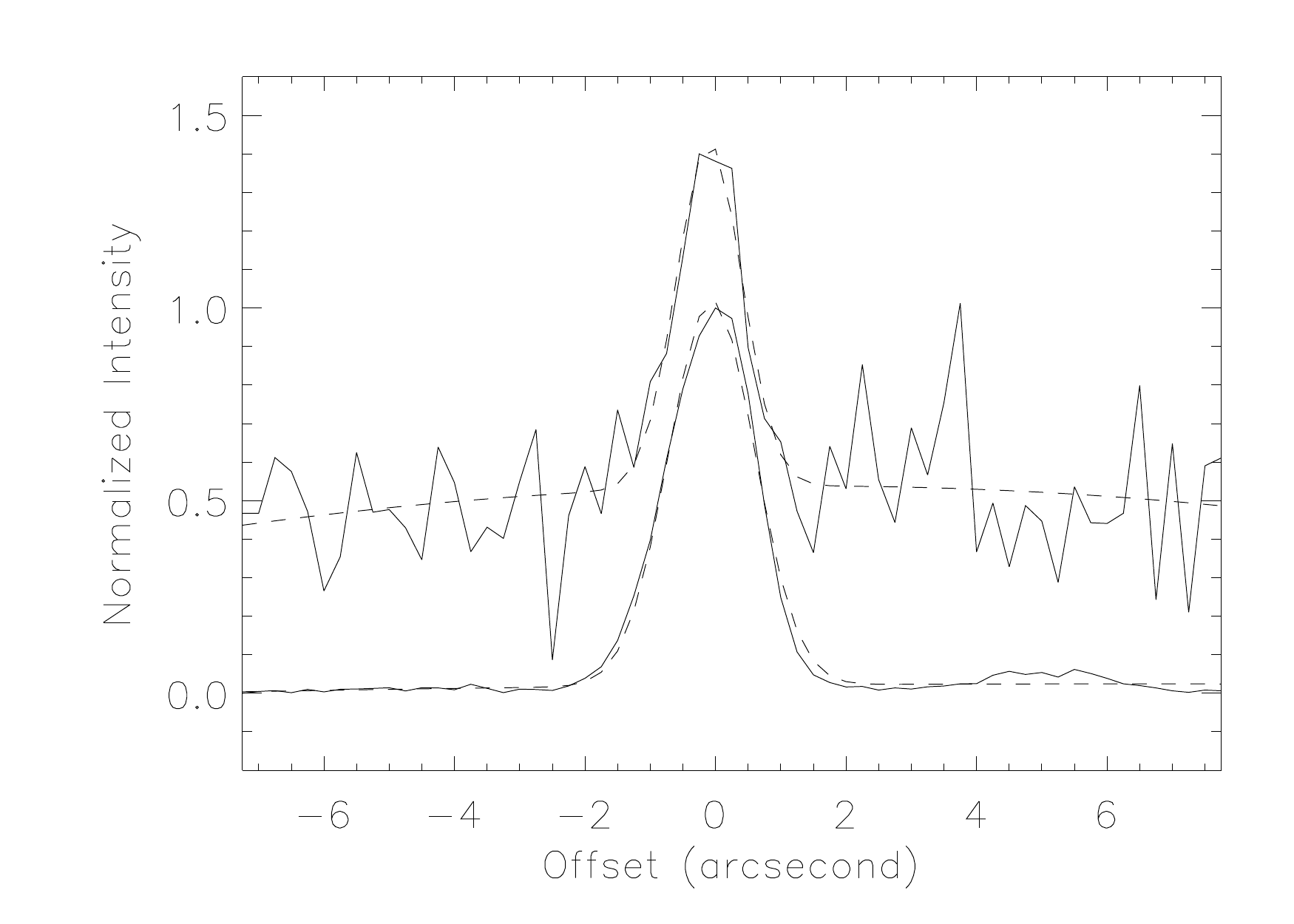} 
\caption{Comparison of the spatial distribution between \heii\ and H$\beta$ lines.
The FWHM of the \heii\ line is $\simeq$ 1\farcs{2},
while that of H$\beta$ is $\simeq$ 1\farcs{6}.}
\label{fig:heii}
\end{figure*}

One notable feature in the spectrum of the source
is the \heii\ ($\lambda$4686 \AA) line which is known to be often produced
by strong X-ray photoionization, including those from 
ULXs (e.g., Moon et al. 2011). 
The presence of the \heii\ emission also supports that it is 
a real optical counterpart to \xmmc.
At the distance of 33.5 Mpc, the de-reddend luminosity of the \heii\ emission is
$\sim$ 1.4 $\times$ 10$^{37}$ ergs s$^{-1}$, corresponding 
to the X-ray (0.1--12 keV) to \heii\ luminosity ratio of $\sim$ 5.7 $\times$ 10$^{4}$. 
This makes \xmmc\ one of the most luminous ULXs observed in the \heii\ emission.
Figure~\ref{fig:heii}, which compares the spatial distribution of the \heii\
emission with that of the nearby H$\beta$ emission, 
shows that the former is a bit more centrally concentrated than the latter --
the FWHM of the \heii\ emission is $\simeq$ 1\farcs{2}, whereas that of H${\beta}$ is $\simeq$ 1\farcs{6}.
The size of the H$\beta$ emission
is substantially greater than the seeing size ($\sim$ 1\arcsec\ in average)
and also the size of the \heii\ emission. 
The size  of the \heii\ emission, however, is not much different from the seeing size,
and it is uncertain if the former is indeed greater than the latter.
If the large $\simeq$ 410 \kms\ velocity dispersion of the \heii\ emission (\S2)
is due to the rotation of a photoionized accretion disc around the X-ray source
as suggested for other ULXs \citep[see][]{met11}, 
the size of an area where the majority of the \heii\ emission originates 
is $\sim$ $GM/v^2$, where $G$, $M$ and $v$ are the Gravitational constant, 
mass and velocity dispersion, respectively. 
This gives $\simeq$ 0.05--5 AU for the mass range of 10--1,000 \msun, 
which is in fact much smaller than the seeing size.
Note that the H$\beta$ emission dose not show any apparent velocity dispersion (\S2),
and it is in a different dynamic motion from \heii.
We thus conclude the linear size of the H$\beta$ emission to be 260~pc corresponding to 1\farcs{6}, 
while the \heii\ emission is unresolved.
This is consistent with what have been observed in other ULXs where
large ($\simgt$ 100 pc) emission nebulae show more extended emission 
in H$\beta$ than \heii\ \citep{met11}.
In particular, the ULX Ho IX X-1 at 3.6 Mpc shows an unresolved central \heii\ emission of $\simeq$ 370 \kms\ 
velocity dispersion surrounded by diffuse \heii\ emission as well as by strong unevenly-distributed H$\beta$ emission.
At an about 10 times larger distance, the \heii\ emission of Ho IX X-1 may appear as that of \xmmc\ presented in this study.

\xmmc\ is one of the most luminous ULXs with its X-ray luminosities reaches up to 
$\sim$ 1.5 $\times$ 10$^{41}$ \ergs.
Using the SDSS data, we discovered its optical countepart near an \hii\ region in a spiral arm
of the host galaxy NGC~470, and our optical spectrum of the source shows blue continuum embedded
with numerous emission lines such as H recombinations, forbidden transitions, and most notably
high-ionization \heii\ line at the same velocity of the host galaxy,
validating the high X-ray luminosity of the ULX and confirming its optical conterpart.
The absence of any significant dynamic motion, other than the large velocity dispersion 
of the \heii\ emission, 
of the source identified in the optical spectrum indicates that it is less likely
associated with disruption or merge of satellite galaxies or motion of a supermassive black hole.
The unresolved \heii\ emission of $\simeq$ 410 \kms\ velocity dispersion,
which is consistent with what have been observed in other ULXs, 
most likely represents a compact ($<$ 5 AU) photoionized accretion disc
rotating the central X-ray source.
The large optical luminosity suggests that the source may be an unresolved stellar cluster.

\acknowledgments 
Based on observations made with the Gran Telescopio Canarias (GTC) instaled in the Spanish
Observatorio del Roque de los Muchachos of the Instituto de Astrof\'\i sica de Canarias, in the island of La Palma. We have used the online databases: Sloan Digital Sky Survey (http://www.sdss.org/) and NED (NASA Extragalactic Database,
http://nedwww.ipac.caltech.edu/).

\clearpage

\newpage

\newpage

\begin{deluxetable}{cccc||cccc}
\tabletypesize{\footnotesize}
\tablecolumns{8}
\tablewidth{0pt}
\tablecaption{List of observed lines\label{tbl_line}}
\tablehead{
\colhead{Line} & \colhead{$\lambda$}  & \colhead{Intensity}   & \colhead{FWHM}  & \colhead{Line}     & \colhead{$\lambda$}  & \colhead{Intensity}   & \colhead{FWHM}  \\
\colhead{}     & \colhead{(\AA)}      & \colhead{}            & \colhead{(\AA)} & \colhead{}         & \colhead{(\AA)}      & \colhead{}            & \colhead{(\AA)}
}
\startdata
\oii            &  3728.2             &  643.0(16.6)          & 6.1(0.2)            & H$\eta$            &  3835.6              & 10.6(1.0)             & 
4.5(0.8)         \\ 
\neiii          &  3868.8             &  39.4(1.5)            & 4.9(0.6)            & \hei               &  3889.2              & 26.0(1.2)             & 
5.3(0.7)           \\ 
\neiii+\hei     &  3968.9             &  32.0(1.1)            & 6.4(0.4)            & H$\delta$          &  4102.1              & 29.1(0.9)             & 
5.6(0.5)           \\ 
H$\gamma$       &  4340.7             &  45.8(1.2)            & 5.9(0.2)            & \oiii              &  4362.5              & 3.3(0.3)              & 
5.9(0.4)           \\ 
\hei            &  4471.3             &  3.2(0.2)             & 5.0(0.3)            & \heii              &  4685.4              & 4.6(0.2)              & 
9.0(0.6)           \\ 
H$\beta$        &  4861.3             &  91.6(2.1)            & 6.6(0.2)             & \oiii              &  4958.7              & 59.6(1.4)             & 
6.8(0.2)           \\ 
\oiii           &  5006.7             &  177.0(4.0)           & 7.0(0.1)            & \nitrogeni         &  5199.0              & 6.1(0.2)              & 
7.3(0.4)           \\ 
\nii            &  5755.4             &  1.7(0.1)             & 5.7(0.3)            & \hei               &  5875.4              & 10.6(0.3)             & 
7.2(0.3)          \\ 
\oi             &  6299.6             &  28.7(0.7)            & 7.1(0.3)            & \oi                &  6363.0              & 9.6(0.3)              & 
6.6(0.4)          \\ 
\nii            &  6547.7             &  40.8(0.8)            & 7.0(0.2)            & H$\alpha$          &  6562.4              & 308.0(6.6)            & 
7.1(0.1)           \\ 
\nii            &  6583.0             &  117.0(2.4)           & 7.1(0.1)             & \hei               &  6677.6              & 4.5(0.2)              & 
7.8(0.5)           \\ 
\sii            &  6715.9             &  72.4(1.6)            & 6.8(0.1)             & \sii               &  6730.3              & 51.8(1.1)             & 
6.7(0.2)           \\ 
\hei\tablenotemark{\dag}       &  7065.8             &  4.2(0.2)             & 11.2(0.8)            & \oiii+\oii         &  7079.3              & 8.8(0.3)    
         &  11.2(0.6)          \\ 
\ariii          &  7135.1             &  9.0(0.3)             & 6.3(0.5)             & \oii               &  7318.9              & 6.1(0.2)              & 
7.3(0.9)           \\ 
\oii            &  7329.4             &  4.0(0.2)             & 5.3(0.8)             & \nodata            &  \nodata             & \nodata               &  \nodata       \\ 
\enddata
\tablenotetext{\dag}{Contaminated by an unidentified nearby feature.}
\tablecomments{The wavelengths are rest-framed observed values and 
the line intensity is in the unit of 10$^{-17}$ ergs s$^{-1}$ cm$^{-2}$.
The values in the parentheses represent 1 $\sigma$ uncertainty level.
}
\end{deluxetable}

\end{document}